\pdfoutput=1

\documentclass[11pt]{article}

\usepackage[]{acl}
\usepackage{times}
\usepackage{latexsym}

\usepackage[T1]{fontenc}

\usepackage[utf8]{inputenc}

\usepackage{microtype}

\usepackage{inconsolata}

\usepackage{graphicx}

\usepackage{enumitem}
\usepackage{caption}
\usepackage{amsmath}
\usepackage{multirow}
\usepackage{breqn}
\usepackage{subcaption} 
\usepackage{float}
\usepackage{adjustbox}
\usepackage{booktabs}
\usepackage{colortbl}
\usepackage{xcolor}

\title{Tailoring Table Retrieval from a Field-aware Hybrid Matching Perspective}

\author{Da Li \textsuperscript{\rm{1,2}}\space\space
Keping Bi\textsuperscript{\rm{1,2}}\space\space
Jiafeng Guo\textsuperscript{\rm{1,2}}\space\space
Xueqi Cheng\textsuperscript{\rm{1,2}}\\
    \textsuperscript{\rm 1}CAS Key Lab of Network Data Science and Technology, ICT, CAS\\
    \textsuperscript{\rm 2}University of Chinese Academy of Sciences\\
    \{lida21s, bikeping, guojiafeng, cxq\}@ict.ac.cn
}

\begin{document}
\maketitle

\newcommand{\modelname}{THYME}
\newcommand{\mean}{\underset}
\begin{abstract}
\textbf{Table retrieval}, essential for accessing information through tabular data, is less explored compared to text retrieval. The row/column structure and distinct fields of tables (including titles, headers, and cells) present unique challenges. For example, different table fields have varying matching preferences: cells may favor finer-grained (word/phrase level) matching over broader (sentence/passage level) matching due to their fragmented and detailed nature, unlike titles. This necessitates a table-specific retriever to accommodate the various matching needs of each table field. Therefore, we introduce a \textbf{T}able-tailored \textbf{HY}brid \textbf{M}atching r\textbf{E}triever (\textbf{\modelname}), which approaches table retrieval from a field-aware hybrid matching perspective. Empirical results on two table retrieval benchmarks, NQ-TABLES and OTT-QA, show that THYME significantly outperforms state-of-the-art baselines. Comprehensive analyses confirm the differing matching preferences across table fields and validate the design of THYME.
\end{abstract}

\section{Introduction}

Table retrieval is an important way for seeking information stored in tables, organized in rows and columns \citep{cafarella2008webtables,jauhar2016tables,zhang2020web}. For example, in the Natural Questions dataset constructed from Google queries targeting Wikipedia pages, information needs requiring tables account for 25.6\% of all questions \citep{kwiatkowski-etal-2019-natural}. Relevant tables serve as foundations for table-related tasks such as question answering \citep{cafarella2008webtables,jauhar2016tables} and fact verification \citep{chen2020tabfact}, which may involve converting natural language questions to SQL. Despite extensive studies on unstructured text retrieval, structured table retrieval remains under-explored. We aim to enhance table retrieval performance to better serve table-related information needs.

\begin{figure}[tbp]
    \centering
    \includegraphics[width=\linewidth]{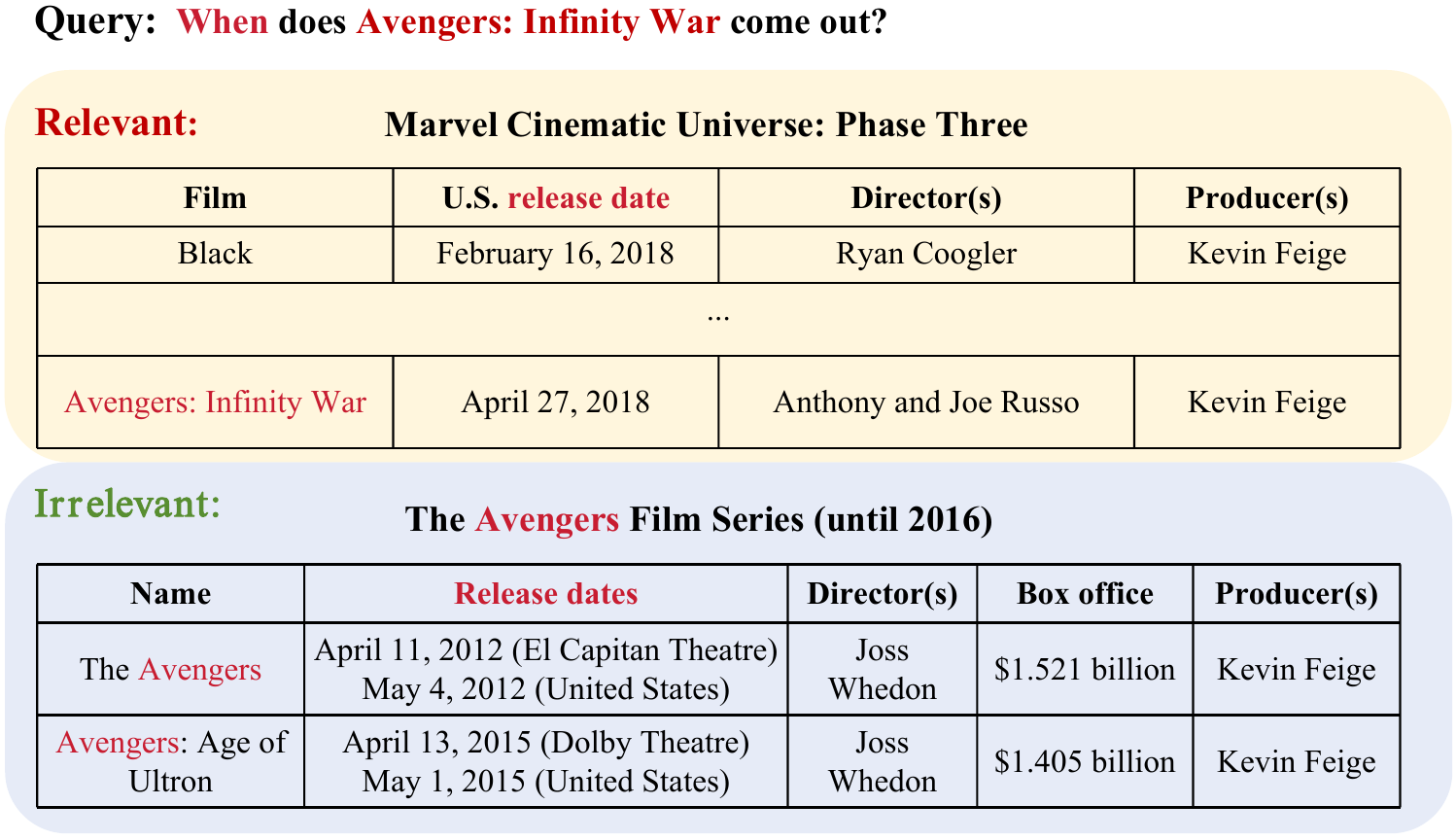}
    \caption{A case of table retrieval showing fine-grained matching in cells is important. 
    }
    \label{fig:match_pattern}
\end{figure}

Table retrieval differs from text retrieval in several ways: 1) Text data are usually unstructured, while tables have a structured format with rows, columns, headers, and cells, suggesting a table-specific encoding approach. 2) Unlike documents whose sentence order matters, tables' data entry order does not affect their information. 3) Each row in a table is equally important, making it challenging to compress table information into dense representations. 4) Table cells contain detailed information, often in words or phrases, making local finer-grained matching more critical than in document retrieval. Figure~\ref{fig:match_pattern} shows an example to illustrate the importance of local lexical matching in table retrieval.

Existing table retrieval methods have explored encoding techniques to fit the row and column structure~\citep{Herzig_2020,yin2020tabert}, condensing table information by selecting cells~\citep{Herzig_2020} or aggregating rows~\citep{Trabelsi_2022}, and pre-training tasks to facilitate table retrieval~\citep{herzig2021open,chen2023bridge}. However, previous methods often compress tables into dense vectors, which may not capture parallel data rows and fine-grained cell matching effectively. Thus, we aim to improve table retrieval by focusing on these aspects.

We approach it from a field-aware hybrid matching perspective, incorporating both sparse and dense representations. Sparse representations \cite{formal2021splade} can retain detailed information as tokens, complementing the global semantics carried in dense representations, which suits table cells better. Dense representations may be better for titles or captions, similar to passages \cite{guo2025came} as unstructured texts. 
A challenging issue is how to adaptively learn the suitable representation and matching pattern for each table field. 

To this end, we introduce THYME, a \textbf{T}able-tailored \textbf{HY}brid \textbf{M}atching r\textbf{E}triever for field-aware hybrid matching. Using a shared encoder, we construct dense and sparse representations for queries and tables. 
The $[CLS]$ embedding implicitly captures field importance/preference on coarse-grain semantics through extensive Transformer interactions.  
In contrast to dense vectors, sparse representations hold greater potential for field-specific considerations due to their explicit segmentation of content tokens by field. Therefore, we focus on learning field-aware sparse representations, which can reflect field importance on fine-grained semantics. 
Based on a shared encoder, sparse and dense representations can be learned coordinately and finalize the field suitability for coarse and fine grains of semantics during relevance matching.  
Specifically, for sparse table body (headers and cells) representations, we leverage mean pooling to retain similar types of information within columns and max pooling to capture more important semantics across columns. 
Then, we aggregate the sparse representation of each field (title, header, cell) with a mixture-of-expert mechanism, dynamically retaining the term information relevant to the matching.
The final relevance score sums up the dense and sparse matching scores, with a score dropout strategy during training to adaptively learn from the lexical and semantic matching paths.

We evaluate THYME on table retrieval benchmarks, NQ-TABLES~\citep{herzig2021open} and OTT-QA~\citep{Chen_Chang_Schlinger_Wang_Cohen_2020}, showing that it outperforms state-of-the-art baselines, including sparse, dense, and hybrid retrievers. Analyses confirm that table titles prefer dense matching, while headers and cells prefer sparse matching, and THYME effectively captures these preferences. Within the RAG framework, THYME enhances the results of various LLMs by providing more relevant and accurate tables. Our studies indicate a promising way of elevating table retrieval, which can shed light on future research on this topic.
\section{Related Work}
\subsection{Text Retrieval}
Typical text retrievers can be classified into three types based on representations used: sparse, dense, and hybrid retrieval which combines them.

Sparse retrievers refer to models that use sparse representations such as TF-IDF~\citep{sparck1972statistical} and BM25~\citep{robertson1994some}. Building on pre-trained language models (PLMs), SparTerm~\citep{bai2020sparterm} and SPLADE~\citep{formal2021splade} aim to generate term distributions over vocabulary. Dense retrievers usually map inputs to continuous vectors based on PLMs~\citep{devlin2019bert,liu2019roberta}. The hidden state of $[CLS]$ as the representation of the query and passages are used for matching. 

Dense and Sparse retrieval have distinct advantages. While dense retrieval generally outperforms sparse retrieval, the latter can be more effective in cold-start situations or when relevant data is limited. Hybrid retrieval can combine the advantages of them~\citep{craswell2020overview,bajaj2018ms}. A straightforward approach is to train two different types of retrievers independently and then combine their outputs linearly to give a final relevance score~\citep{DBLP:journals/corr/abs-2110-06918, DBLP:journals/corr/abs-2010-01195, DBLP:journals/corr/abs-2112-04666, DBLP:journals/corr/abs-2005-00181, guo2025came}. Inspired by the boosting technique, CLEAR~\citep{DBLP:journals/corr/abs-2004-13969} tries to learn a BERT-based retriever from the residual of BM25. In addition, knowledge distillation is also an approach to realize the hybrid retrieval~\citep{shen2023unifier}.

\subsection{Table Retrieval}
Table retrieval, as introduced by \citet{Zhang_2018}, entails extracting tables from a large corpus and aligning them with specific queries. They also proposed a Semantic Table Retrieval system that leverages features between queries and tables to facilitate retrieval.

Since PLMs are primarily trained on textual corpora, they cannot comprehend structured data like tables. To improve the comprehension of tables, TAPAS~\citep{Herzig_2020} utilizes distinct types of embeddings like row and column embeddings, to represent the row and column structure of the table based on BERT~\citep{devlin2019bert}. In addition to specialized network design, another approach involves training on a specific table corpus to enhance the pre-trained model’s understanding of tables. UTP~\citep{chen2023bridge} incorporates the table and its surrounding text to develop an unsupervised alignment loss function, prompting the model to learn a unified representation across table-text modalities. 
Given the complexity of table structure and content, a single dense representation may not fully capture the table's details, SSDR~\citep{jin2023enhancing} aims to represent both the query and table through multiple vector representations that integrate syntactic and structural information. This helps to extract information meticulously from the query and table for matching. Alongside single-table retrieval, multi-table retrieval has also garnered attention from the research community~\citep{chen2024tableretrievalsolvedproblem}.

Given that table retrieval is less explored, it is possible to replicate the methods used in table reranking to enhance retrieval performance. TaBERT~\citep{yin2020tabert} and StruBERT~\citep{Trabelsi_2022} take query-table pairs as input and set up a vertical/horizontal self-attention mechanism between cell representations of different rows and columns. Additionally, other approaches take into account the table layout and structure, representing tables as hypergraphs by defining various types of nodes and edges~\citep{Wang_2021}. These methods indicate the important role of structure in table representation.
\section{Task Description}
Let $D=\{(q_i,T_i^+)\}_{i=1}^N$ be a labeled dataset, where $q_i$ denotes an individual query and $T_i^+$ is a set of tables $\{t_i^+\}$ that are considered relevant to $q_i$. The number of elements in $T_i^+$ varies depending on the query. Table retrieval aims to train a retriever that can identify relevance matching patterns of a query $q_i$ and its relevant tables $T_i^+$, considering each table's specific structure and fields. After training, this retriever is expected to accurately capture the nuances of relevance by understanding how the $title$, $headers$, and $cells$ fields align with the information needs expressed in the query. Ultimately, for any new query $q$, the retriever can retrieve the relevant tables from a collection of tables.

\section{Table-Tailored Hybrid Matching Retriever (THYME)}

\begin{figure*}[htbp!]
    \centering
    \includegraphics[width=0.95\linewidth]{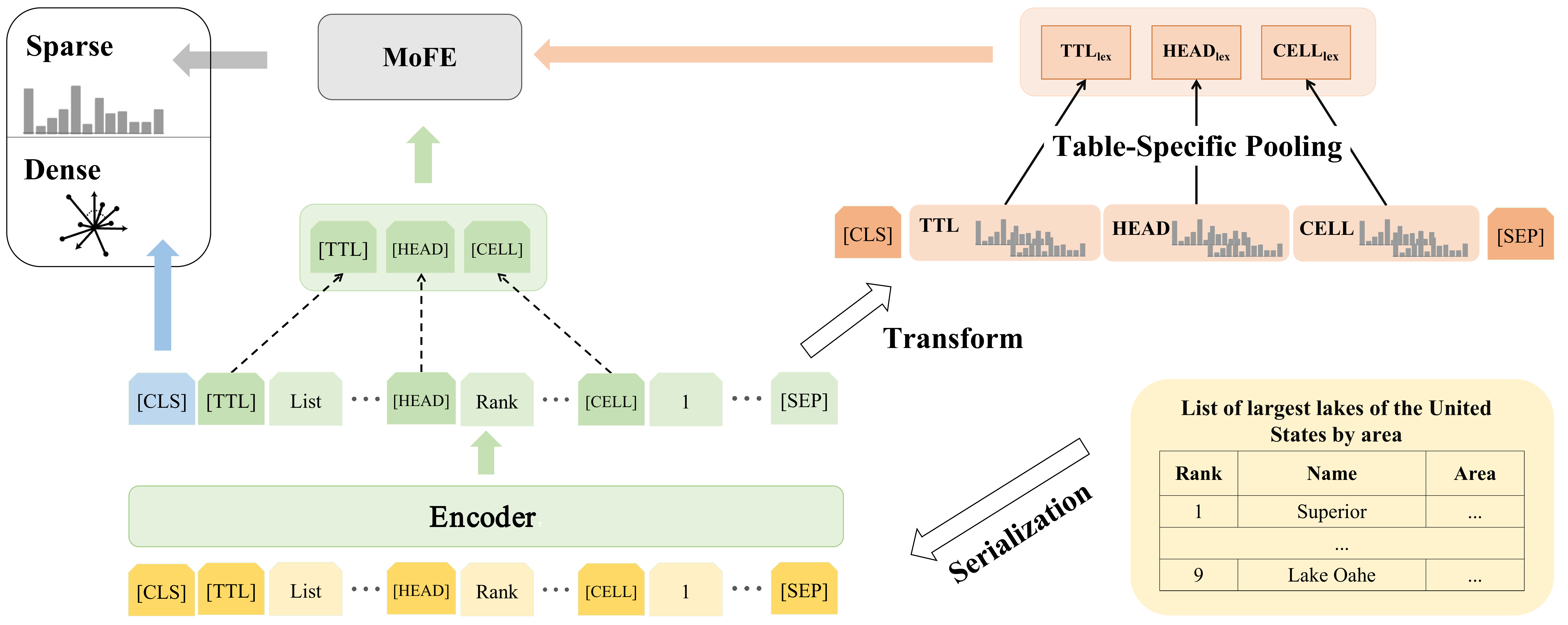}
    \caption{Illustration of Table-Tailored Hybrid Matching. Serialized tables are encoded through an encoder shared with queries. }
    \label{fig:model_architecture}
\end{figure*} 

In this section, we introduce \modelname, a \textbf{T}able-tailored \textbf{HY}brid \textbf{M}atching r\textbf{E}triever. Figure~\ref{fig:model_architecture} shows the overall model architecture. Dense representations capture semantics through extensive Transformer interactions, which are effective for extracting information from unstructured text, such as titles. Sparse representations can retain detailed information and are suitable for cell-specific information needs. To better capture the appropriate relevance matching patterns (lexical, semantic, and at various granularities) for different table fields, we incorporate dense and sparse representations, propose a field-aware lexical matching mechanism, and craft a hybrid training strategy. Note that we employ BIBERT ~\citep{bibert-pretrain} and SPLADE ~\citep{formal2021splade} as the backbones to calculate dense and sparse representations. Note that other advanced backbones can be alternatives to achieve better performance but our focus is to study table-specific hybrid matching. 
Next, we detail each component of THYME.

\subsection{Query Representation}
Since query $q$ is unstructured text without special processing, we directly obtain directly its dense and sparse representation based on BIBERT \cite{bibert-pretrain} and SPLADE~\citep{formal2021splade} respectively. The hidden state of $[CLS]$ is represented as the dense representation. 
The sparse representation is obtained by applying max pooling over the entire sequence:
\begin{equation}
\resizebox{0.75\linewidth}{!}{$
\begin{aligned}
&\mathbf{H}_{q} = \text{Enc}(q), \, \mathbf{W}_q = \text{trans}(\mathbf{H}_{q}), \\
&\mathbf{q}_{cls} = \mathbf{H}_{q}[CLS], \\
&\mathbf{q}_{lex} = \max_{i \in |q|} \log(1+\text{ReLU}(\mathbf{W}_{q}[i])),
\end{aligned}
$}
\end{equation} where $\mathbf{q}_{cls}$ and $\mathbf{q}_{lex}$ correspond to dense and sparse representations, respectively, $trans(\cdot)$ is a linear used to map the output of $Enc(\cdot)$ to the distribution in the vocabulary space.

\subsection{Table Serialization}
To encode tables with Pre-trained Language Models (PLMs) and indicate the title, headers, and cells of a table $t$, we insert the special tokens $[TTL]$, $[HEAD]$, and $[CELL]$ before the corresponding tokens in the fields. For a table $t$ with $title$, n headers - $headers_n$, and $cell_{m\times n}$ of m rows and n columns, it can be serialized as: 
\begin{equation}
\resizebox{0.9\linewidth}{!}{$
\begin{aligned}
t  &= [CLS], [TTL], title, [HEAD], header_{0}, \\\nonumber 
   &\ldots header_{n-1}, [CELL], cell_{0,0}, cell_{0,1},\ldots, \\
   & cell_{m-1,n-1}, [SEP]. 
\end{aligned}
$}
\end{equation}
The sequence will be fed into the encoder to construct hybrid representations.

\subsection{Global Semantic Matching}
The global semantics of a table are essential to satisfy topic-related queries. The semantics represent the global information of the table and need to converge all the information of the fields in it. The self-attention mechanism can help the $[CLS]$ embedding aggregate the information stored in each field differently, and the field indicator tokens in the encoding sequence can help $[CLS]$ to identify the boundary of the fields and thus produce better weights. The dense representation of table $t$ is generated from: 
\begin{equation}    
\begin{aligned}
\mathbf{t}_{cls} &= \text{Enc}(t)[CLS].
\end{aligned}
\end{equation}
We have also tried the pooling strategy similar to that of sparse retrieval to obtain the dense representation. However, it does not perform better than simply using $[CLS]$, indicating that the $[CLS]$ can effectively differentiate the field importance. The semantic matching scores between the query and the table are obtained in the following way:

\begin{equation}    
\begin{aligned}
s_{sem}(q,t) &= \text{sim}(\mathbf{q}_{cls},\mathbf{t}_{cls}).
\end{aligned}
\end{equation}
We choose the inner product as the similarity function.

\subsection{Field-aware Lexical Matching}
The body of the table, including headers and cells, mainly consists of words or phrases that lack coherent semantics. Compared to dense vectors, sparse representations are more effective in distinguishing across different fields of a table. They explicitly aggregate content by fields while retaining fine-grained distinctions. We hope to differentiate different fields by sparse representations and obtain the final sparse representations of tables with differentiated emphasis on these fields. 
To this end, we propose a table-specific pooling mechanism and conduct field-level aggregation to facilitate field-aware lexical matching.

\textbf{Table-Specific Pooling.}
First, to obtain sparse representations, the formatted input of table $t$ is transformed into a sequence of logits $\mathbf{W}_{t}\in \mathcal{R}^{|t|*|V|}$: 
\begin{align}
\label{eq:lexical_table}
\mathbf{H}_{t} &= \text{Enc}(t),\mathbf{W}_{t}= \text{trans}(\mathbf{H}_{t}).
\end{align}
Then, we adopt different pooling methods for the table fields. Max pooling is good at capturing fine-grained information within a sequence while mean pooling emphasizes every piece of information in the sequence. Based on this presumption, for titles, headers, and cells, our pooling strategies are as follows:

\textit{Titles}:
Since table titles are unstructured text, similar to queries, we use max pooling as in \cite{formal2021splade}: 
\begin{equation}    
\resizebox{0.85\linewidth}{!}{$
\begin{aligned}
\label{eq:sparse_title}
\mathbf{title}_{lex} &= \max_{token_i \in {title}}\, log\,(1+\text{ReLU}(\mathbf{W}_t[i])\,),
\end{aligned}
$}
\end{equation}
where $i$ represents the index of the token in the title.

\textit{Headers:} Headers reflect the relationship embedded in the table. We use mean pooling for the tokens within each header and max pooling across headers to construct the sparse vectors of headers:
\begin{equation}    
\resizebox{0.85\linewidth}{!}{$
\label{eq:sparse_header}
\begin{aligned}
&\mathbf{header}^j_{lex} = \mean{token_i \in {header^j}}{\text{mean}}\, log\,(1+\text{ReLU}(\mathbf{W}_t[i])\,),\\
&\mathbf{headers}_{lex} = \max_{1\leq j\leq n}\, \mathbf{header}^j_{lex},
\end{aligned}
$}
\end{equation}
where $header^j$ is one of the $n$ headers in table $t$. 

\textit{Cells:}
Cells in the same column have identical properties indicated by the corresponding header. The semantics carried in each cell within a column are equally important to represent the column, so we first use mean pooling on the tokens of the same column to obtain column representations. In contrast, different columns are of different importance during matching. Hence, we adopt max pooling over the column representations to retain the essential table information and yield the final lexical representations. In concrete, the sparse vectors of cells are calculated as:
\begin{equation}    
\resizebox{0.85\linewidth}{!}{$
\label{eq:sparse_cell}
\begin{aligned}
&\mathbf{col}^j_{lex} = \mean{token_i \in col^j}{\text{mean}}\, log\,(1+\text{ReLU}(\mathbf{W_t}[i])\,),\\
&\mathbf{cells}_{lex} = \max_{1\leq j \leq n}\, \mathbf{col}^j_{lex},
\end{aligned}
$}
\end{equation}
where $col^j=cell_{0,j}, cell_{1,j}, \cdots, cell_{m,j}$ is the $j$-th column of $t$ which have $m$ cells. $i$ is the index of the token in $col^j$. 

\textbf{Mixture of Field Experts (MoFE).} The information needs of users are varied, ranging from global matching to local matching. Different fields sometimes need to be integrated, while at other times, trade-offs must be made to highlight the most relevant matching signals. 
To adaptively and dynamically aggregate information from different fields for the final representation, we conduct a Mixture of Field Experts (MoFE) to integrate different field representations adaptively. 
Specifically, we select the hidden states of $[TTL]$, $[HEAD]$, and $[CELL]$ to evaluate the importance of sparse representations corresponding to different fields for matching. Based on the field representations yielded from Equations \eqref{eq:sparse_title}, \eqref{eq:sparse_header}, and \eqref{eq:sparse_cell}, the final sparse representation $t_{lex}$ is generated according to:
\begin{equation}
\resizebox{0.85\linewidth}{!}{$
\begin{aligned}
&\mathbf{t}_{gate} = [\,\mathbf{H}_{t}{[TTL]},\mathbf{H}_{t}{[HEAD]},\mathbf{H}_{t}{[CELL]}\,],\\
&\mathbf{t}_{field} = [\,\mathbf{title}_{lex}, \mathbf{headers}_{lex},\mathbf{cells}_{lex}\,], \\
&\mathbf{g}_{field} = \text{softmax}(\,\text{linear}(\,\mathbf{t}_{gate}\,)\,),\\
&\mathbf{t}_{lex} = \sum_{1}^{k} \mathbf{g}_{field}[i]*\mathbf{t}_{field}[i], \, \text{where}\,1 \leq k \leq 3, \\
\end{aligned}
$}
\end{equation} where $\mathbf{t}_{gate}$ is the list of hidden states of $[TTL]$, $[HEAD]$, and $[CELL]$, $\mathbf{t}_{field}$ is the corresponding sparse representations of different fields, $\mathbf{g}_{field}$ adjusts each field of the inflow representations. To obtain the final sparse representation, we introduce the parameter $k$ to control the utilization of different fields, either through selection or aggregation.
The lexical similarity function calculates the lexical matching score between the query and the table.
\begin{equation}
\begin{aligned}
s_{lex}(q,t) &= \text{sim}(\mathbf{q}_{lex},\mathbf{t}_{lex}).
\end{aligned}
\end{equation}
In this paper, we use the inner product as the similarity function, the same way we use the semantic matching score.

\subsection{Hybrid Training}
To let the retrievers learn field-aware lexical matching and global semantic matching simultaneously and sufficiently, we adopt a dropout strategy:

\textbf{Matching Score Dropout.} We use the semantic matching score $s_{sem}(q,t)$ and the lexical matching score $s_{lex}(q,t)$ as the final relevance matching score with probability $p_{sem}/p_{lex}$ during training. For the rest of the training, the overall relevance score is the sum of the two, i.e.:
\begin{equation}
\resizebox{0.87\linewidth}{!}{$
s(q,t) = 
\begin{cases}
s_{sem}(q,t), & p_{sem},\\
s_{lex}(q,t), & p_{lex},\\
s_{sem}(q,t) + s_{lex}(q,t), & 1-p_{sem}-p_{lex}.
\end{cases}
$}
\end{equation}
In this way, the global semantic matching of dense representations and field-aware lexical matching of sparse representations are learned both separately and jointly, facilitating sufficient individual learning and cooperation. In our experiments, we set $p_{sem}=p_{lex}$ since we consider them to have equal importance for training.

\textbf{Loss Function.} We adopt the commonly in-batch negatives and cross-entropy loss for training~\cite{qu2021rocketqa}. Specifically, for a query $q_i$ in a batch, we pair a positive table $t_i^+$, with a set of random negative tables (positive tables from the other queries in the batch, e.g., $\{t_{i,j}^-\}$ for query $q_j$ in the batch), the relevance loss for this sample is computed as:

\begin{equation}
\resizebox{0.7\linewidth}{!}{$
\ell_{rel} = -log \frac{e^{s(q_i,t_i^+)}}{e^{s(q_i,t_i^+)} +\sum_je^{s(q_i,t_{i,j}^-)}}.
$}
\end{equation}

We also introduce the FLOPS regularization~\citep{paria2020minimizingflopslearnefficient} to improve efficiency during training, the training objective can be defined as follows:

\begin{equation}
\resizebox{\linewidth}{!}{$
\ell_{all} = 
\begin{cases} 
\ell_{rel} + (\lambda_q \ell_{FLOPS}^q + \lambda_t \ell_{FLOPS}^t), & s(q,t)=s_{lex}(q,t), \\
\ell_{rel}, & otherwise,
\end{cases}
$}
\end{equation}
where regularization weights ($\lambda_q$ and $\lambda_t$) for queries and tables respectively allow more pressure on the sparsity for exact matching, which is critical for fast retrieval.

In the inference phase, we sum the semantic and lexical matching scores as the final relevance score:
\begin{equation}
\begin{aligned}
s(q,t) = s_{sem}(q,t) + s_{lex}(q,t).
\end{aligned}
\end{equation}
\section{Experimental Settings}
\subsection{Datasets}
We conduct experiments on two standard table retrieval benchmarks:
\begin{itemize}[leftmargin=*]
    \item \textbf{NQ-TABLES~\citep{herzig2021open}} is a subset of the Natural Questions (NQ)~\citep{kwiatkowski-etal-2019-natural}, which is collected from Wikipedia and contains the tables and their corresponding questions.
    \item \textbf{OTT-QA~\citep{kostić2021multimodal}} is an open-domain multi-hop QA dataset of texts and tables from Wikipedia. We use the subset relevant to the table for evaluation.
\end{itemize}
The statistics of our benchmarks are shown in Table~\ref{table:benchmark_static}.
\begin{table}[htbp!]
\begin{adjustbox}{width=\linewidth}
\renewcommand{\arraystretch}{0.90}
\begin{tabular}{llrrrr} \toprule
                                             &                              & \multicolumn{2}{c}{\textbf{NQ-TABLES}} & \multicolumn{2}{c}{\textbf{OTT-QA}} \\
\cmidrule(lr){3-4} \cmidrule(lr){5-6} 
                                             &                                        & Train          & Test         & Train        & Test        \\ \midrule
\multicolumn{1}{l|}{\multirow{2}{*}{Query}}  & \multicolumn{1}{l|}{Count}             & 9,594           & 919          & 41,469        & 2,214        \\
\multicolumn{1}{l|}{}                        & \multicolumn{1}{l|}{Avg. \# Words.}  & 8.94           & 8.90        & 21.79        & 22.82      \\ \midrule
\multicolumn{1}{l|}{\multirow{3}{*}{Table}} & \multicolumn{1}{l|}{Count}             & 169,898         & 169,898       & 419,183       & 419,183      \\
\multicolumn{1}{l|}{}                        & \multicolumn{1}{l|}{Avg. \# Row.}      & 10.70       & 10.70    & 12.90   & 12.90   \\ 
\multicolumn{1}{l|}{}                        & \multicolumn{1}{l|}{Avg. \# Col.}      & 6.10       & 6.10    & 4.80   & 4.80   \\ \midrule
\multicolumn{2}{l|}{\# Golden Tables per Query}                                                 & 1.00            &1.05          & 1.00         & 1.00         \\ \bottomrule
\end{tabular}
\end{adjustbox}
\caption{\label{table:benchmark_static}
Statistics of the Benchmarks. 
}
\end{table}

\subsection{Baselines}\label{methods}
We compare our method with the following baselines. 1) Sparse Retrievers: \textbf{BM25}~\citep{robertson1994some} and \textbf{SPLADE}~\citep{formal2021splade}. 
2) Dense Retrievers: We selected three types of dense retrievers as baselines. The first category includes general retrievers that use a single representation, such as \textbf{BIBERT~\citep{bibert-pretrain}} and \textbf{DPR~\citep{wang-etal-2022-table}}, which has been trained for text retrieval.
The second category includes widely used table retrievers using a single representation, such as \textbf{TAPAS~\citep{Herzig_2020}} and \textbf{DTR~\citep{herzig2021open}}, designed and trained on tabular data. The third category includes table retrievers that use multiple representations, such as \textbf{SSDR$_{im}$~\citep{jin2023enhancing}}, which extracts multiple vectors to represent both queries and tables, enabling fine-grained matching. 
3) Hybrid Retrievers: We introduced hybrid retrievers to fully exploit the strengths of sparse and dense, such as \textbf{DHR}~\citep{lin2023aggretriever}, along with two of our implementations: \textbf{BIBERT+BM25} to investigate the impact of score fusion and \textbf{BIBERT-SPLADE} to analyze the benefits of training. Details of these baselines are 
shown in the Appendix~\ref{baseline_detail}. 
Additionally, some methods, such as TaBERT~\citep{yin2020tabert} and StruBERT~\citep{Trabelsi_2022}, are not used as our baselines, since they are designed for table ranking and not suitable for retrieval.

\begin{table*}[htbp!]
\centering
\begin{adjustbox}{width=\textwidth}
\renewcommand{\arraystretch}{0.90}
\begin{tabular}{lllllllllll} \toprule
             & \multicolumn{5}{c}{\textbf{NQ-TABLES}}            & \multicolumn{5}{c}{\textbf{OTT-QA}}              \\
\cmidrule(lr){2-6} \cmidrule(lr){7-11}                                                              
             & NDCG@5 &NDCG@10 & R@1 & R@10 & R@50 & NDCG@5 & NDCG@10 & R@1 & R@10 & R@50 \\ \midrule
BM25         & 25.52   &27.12    & 18.49  & 36.94     & 52.61     &35.09   &37.45  &23.98   &51.94     & 69.11   \\
SPLADE       & 53.70   &56.75    & 39.84  & 83.33     & 94.65     & 75.27   &76.72  & 62.74   & 89.52     & 95.21      \\ \midrule
BIBERT       & 60.49   &63.16  & 43.78  & 82.25       & 93.71     & 70.57   &72.49  & 56.82   & 86.50     & 94.26      \\
DPR          &63.05   &66.13    &45.32  &85.84 &95.44  &67.92 &70.00   &53.43   &85.95   &93.22 \\
TAPAS        &61.77   &64.29    &43.79  & 83.49       & 95.10     &70.89   &72.72   &57.86   &86.77    &94.04      \\
DTR          &51.04   &53.98    &32.62   &75.86      &89.77       &56.68    &58.94  &42.10   &75.75    &88.80       \\
SSDR$_{im}$  &62.31   &65.02    &45.47   &84.00      &95.05      &69.81  &71.76  &56.96   &86.22     &93.55       \\ \midrule
BIBERT+BM25    &53.81    &57.19    &35.87   &79.63      &94.56      &71.36    &73.29  &59.49   &86.81      &94.67       \\
DHR          &61.16    &64.32    &43.67   &84.65      &\underline{95.62}      &75.27    &76.65   & 63.64   &88.48     &95.30       \\
BIBERT-SPLADE  &\underline{63.24}    &\underline{66.25}    &\underline{45.62}   &\textbf{86.72}      &\underline{95.62}      &\underline{76.90}    &\underline{78.30}  & \underline{64.72}   &\underline{91.01}      &\textbf{96.34} \\ 
 \midrule
\textbf{THYME}   &\textbf{65.72}$^{\dagger}$    &\textbf{68.14}$^{\dagger}$ &\textbf{48.55}$^{\dagger}$   &\underline{86.38}    &\textbf{96.08}     &\textbf{78.21}$^{\dagger}$  &\textbf{79.58}$^{\dagger}$   &\textbf{66.67}$^{\dagger}$    &\textbf{91.10}  &\underline{96.16}        \\ 

\bottomrule
\end{tabular}
\end{adjustbox}
\caption{\label{table:main_result}
Overall table retrieval performance. \textbf{Bold} and \underline{underline} indicate the best and suboptimal performance respectively. Statistically significant (p < 0.05) improvements over BIBERT-SPLADE are marked with $\dagger$.}
\end{table*}

\subsection{Evaluation Metrics}
We use recall and normalized discounted cumulative gain (NDCG) for evaluation. Since table retrieval results will be used for downstream tasks (e.g., table comprehension, tableQA, etc.), considering the efficiency of downstream applications, we use a cutoff at 50 for the retrieved tables, as in~\citep{Herzig_2020,jin2023enhancing} add references to other papers that also evaluate top 50 results. We report Recall@1, Recall@10, and Recall@50 to show how many relevant tables are retrieved, and NDCG@5 and NDCG@10 to show whether target tables are ranked to top positions. Statistical significance is measured with two-tailed t-tests with $p<0.05$. 

\subsection{Implementation Details}
We initialize our model, BIBERT, and SPLADE with BERT-base. For the other baselines, we use the released checkpoints for initialization.
To ensure fair comparisons, we set the batch size, learning rate, and training steps of all the models that we have trained to the same. With a batch size of 144 and a learning rate of $1e-5$, we compare the performance of the different methods after 50 epochs of training. For THYME, we set $p_{sem}=p_{lex}=0.15$ for matching score dropout and $\lambda_q=\lambda_t=1e-4$ for FLOPS regularization.
During the inference phase, we use Faiss~\citep{2022ascl.soft10024J} for ANN search on dense representations and inverted indexes for sparse representations.

\section{Results and Discussion}
In this section, we discuss the results of our experiments. We first
compare the overall performance of THYME with all the baselines described in~\ref{methods}. Then, we study the effect of different model components in the following subsections.

\subsection{Overall Retrieval Performance}
Table \ref{table:main_result} shows the performance of three groups of methods, i.e., sparse, dense, and hybrid retrievers, on NQ-TABLES and OTT-QA. It shows that hybrid retrievers perform better than dense and sparse retrievers. Even the simple combination of BIBERT and SPLADE boosts the performance by a large margin. Among all the methods, our hybrid retriever THYME performs the best, significantly better than the SOTA baselines.

We also have the following observations: 1) By using special tokens in the input, BIBERT achieves comparable results to TAPAS. This suggests that the model can adaptively learn the structure of the table and the special table-specific layers for tables may be not necessary for retrieval. 2) For NQ-TABLES, dense retrievers outperform sparse retrievers, whereas sparse retrievers may be more effective for OTT-QA. This is because queries in OTT-QA contain more detailed information. Dense retrievers are considered suited for semantic matching rather than exact matching, highlighting the importance of hybrid matching in table retrieval. 3) Despite the differences in benchmark characteristics, SSDR$_{im}$ achieves the best average performance among dense retrievers. This suggests that interactive matching models perform better than one-shot matching between two vectors, aligning with findings in text retrieval.

\subsection{Analyses on Model Variants}
\begin{table}[htbp!]
\centering
\begin{adjustbox}{width=\linewidth}
\renewcommand{\arraystretch}{0.90}
\begin{tabular}{lllll} \toprule
\textbf{Pooling}         &\textbf{Aggregation}  & \textbf{NDCG@5} & \textbf{NDCG@10} & \textbf{R@10} \\ \midrule
Table-Specific  &MoFE       & \textbf{65.72}   & \textbf{68.14}  & \textbf{86.38}     \\ \midrule
Max  &MoFE      &62.18$^{\star}$   &64.52$^{\star}$   & 85.79    \\
Mean &MoFE      &60.61$^{\star}$   & 63.96$^{\star}$  & 85.09$^{\star}$     \\ \midrule
Max  &Max       & 61.44$^{\star}$   & 64.19$^{\star}$  & 85.42     \\
Mean &Mean      & 57.72$^{\star}$   & 60.96$^{\star}$  & 83.15$^{\star}$     \\ \bottomrule
\end{tabular}
\end{adjustbox}
\caption{\label{table:different_field}
Comparisons of pooling and aggregation methods for sparse representations on NQ-TABLES. `$\star$' indicates statistically significant differences (p<0.05) with THYME (the first row).}
\end{table} 

To obtain the sparse representations of the table, we conduct table-specific pooling within each field and use MoFE to aggregate the term distributions. To see the effect of our table-specific pooling and MoFE, we train and evaluate other variants for these two components. For pooling within the field, we also tried max or mean pooling on all the tokens instead of the table-specific pooling in THYME. For the aggregation over fields, we also attempted max and mean pooling. Notably, using max/mean pooling both within and across fields degrades to treating tables as unstructured text and representing them with SPLADE. Table~\ref{table:different_field} shows how the variants of THYME perform with the revised sparse representations. We can see that 1) max pooling has better performance than mean pooling, consistent with the observations from SPLADE, but both are significantly worse than our table-specific pooling approach;  2) MoFE is better than using max or mean pooling over the fields, indicating field importance in the final sparse representations should be differentiated. 

\subsection{Matching Preferences of Table Fields} 
To see whether table titles and bodies have different relevance matching preferences due to their structures, we train and evaluate several variants of THYME that mask the dense/sparse representations of titles or bodies when aggregating them to obtain the final ones. For instance, when only dense representations of titles are used, their sparse representations are not aggregated to the tables' final sparse vectors. Table~\ref{table:different_pattern_branch} shows how the variants perform. We can observe: 1) Without sparse representations of titles, performance does not drop as much as the other variants; 2) The performance regresses more without sparse representations of table bodies than dense ones; 3) When only titles' sparse representations and bodies' dense representations are used, THYME has the worst performance, confirming that titles prefer semantic matching and bodies (headers and cells) require lexical matching more.
\begin{table}[htbp!]
\begin{adjustbox}{width=\linewidth}
\begin{tabular}{lllll} \toprule

\textbf{Title}        & \textbf{Headers\&Cells} & \textbf{NDCG@5}  & \textbf{NDCG@10} & \textbf{R@10} \\ \midrule
Dense, Sparse                               &Dense, Sparse                                            & \textbf{65.72}   & \textbf{68.14}  & \textbf{86.83}     \\ \midrule
Dense                                      &Dense, Sparse                                             & 63.97$^{\star}$   & 67.18  & 86.72      \\
Dense, Sparse                              &Sparse                                                    & 63.81$^{\star}$    & 66.48$^{\star}$   &85.65$^{\star}$       \\
Dense, Sparse                              &Dense                                                     &57.98$^{\star}$    & 61.14$^{\star}$   &82.48$^{\star}$       \\
Sparse                                     &Dense                                                     & 57.48$^{\star}$    & 60.25$^{\star}$   & 82.15$^{\star}$       \\ \bottomrule
\end{tabular}
\end{adjustbox}
\caption{\label{table:different_pattern_branch}
Study on the impact of representation types of table fields on NQ-TABLES. `$\star$' marks the statistically significant difference (p<0.05) compared to THYME (the first row).
}
\vspace{-2mm}
\end{table}

\section{Application in TableQA}
We conducted an end-to-end evaluation of the TableQA to demonstrate the practical value of \modelname. Mistral~\citep{jiang2023mistral7b}, Llama3~\citep{grattafiori2024llama3herdmodels}, and Qwen-2.5~\citep{qwen2025qwen25technicalreport} are used to conduct retrieval-augmented generation in an end-to-end manner. The results of the evaluation are in the Table~\ref{table: application}.
\begin{table}[htbp!]
\centering
\begin{adjustbox}{width=\linewidth}
\renewcommand{\arraystretch}{0.8}
\begin{tabular}{llrrr} \toprule
\multirow{2}{*}{Retriever}          & \multirow{2}{*}{LLM} & \multicolumn{3}{c}{Accuracy} \\ \cmidrule(lr){3-5}
                                    &                      & n=1     & n=3     & n=5     \\ \midrule
\multirow{3}{*}{BIBERT}             & Mistral              &32.93   &33.46   &35.30    \\
                                    & Llama3               &32.66   &33.16   &34.28    \\
                                    & Qwen-2.5             &34.80   &37.92   &37.01         \\ \midrule   
\multirow{3}{*}{SPLADE}             & Mistral              &29.61   &35.42   &34.95    \\
                                    & Llama3               &32.07   &37.17   &33.88    \\
                                    & Qwen-2.5             &31.90   &35.79   &37.35          \\ \midrule
\multirow{3}{*}{SSDR$_{im}$}        & Mistral              &32.76   &36.95   &36.30    \\
                                    & Llama3               &34.25   &38.14   &37.89    \\
                                    & Qwen-2.5             &33.42   &39.27   &39.66    \\ \midrule                                     
\multirow{3}{*}{BIBERT-SPLADE}      & Mistral              &32.67   &33.59  &35.71    \\
                                    & Llama3               &32.66   &33.66  &33.92         \\
                                    & Qwen-2.5             &33.24   &36.76  &37.02    \\ \midrule               
\multirow{3}{*}{\modelname} & Mistral                     &35.48  &37.59   &37.20    \\
                                    & Llama3              &36.14   &39.16   &39.29    \\
                                    & Qwen-2.5            &\textbf{37.28}   &\textbf{40.28}   &\textbf{41.20}           \\ \bottomrule
\end{tabular}
\end{adjustbox}
\caption{\label{table: application} End-to-end QA Performance of NQ-TABLES, n indicates the number of relevant tables retrieved by various retrievers in the context.}
\vspace{-2mm}
\end{table}

In general, LLMs generate accurate answers when the relevant table is in the input, but may produce incorrect answers if the table appears at the end of the context. A RAG-friendly table retriever requires not only the ability to retrieve relevant tables but also the capability to rank them at the top of the retrieval list. THYME significantly enhances the accuracy of answers generated by LLMs among various form retrievers.

\section{Case Study}
\begin{figure}[htbp!]
  \centering
  \includegraphics[width=\linewidth]{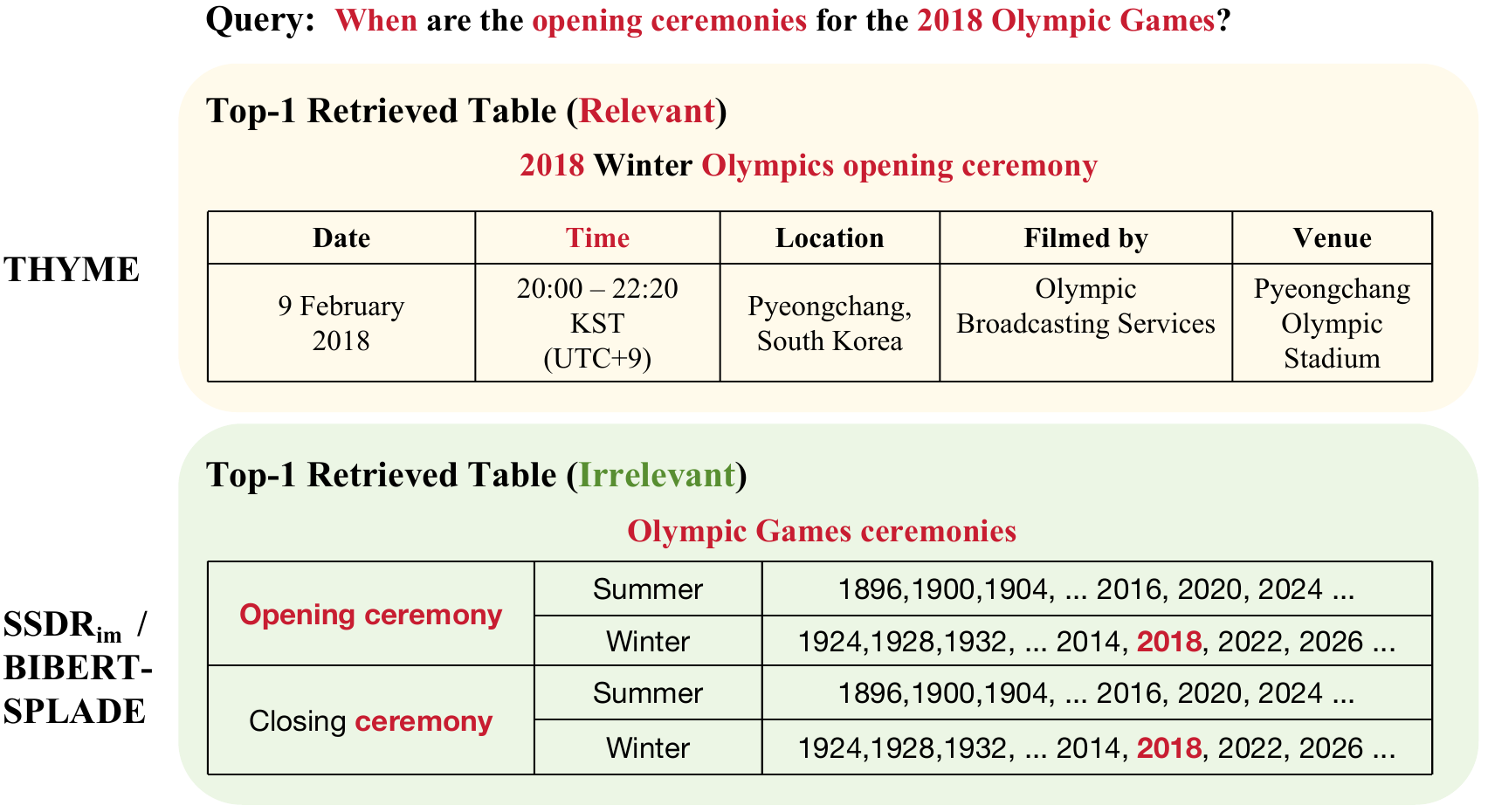}
  \caption{Top-1 retrieved table from different retrievers.}
  \label{fig:case_different_model}
  \vspace{-3mm}
\end{figure}

We show a concrete case in Figure~\ref{fig:case_different_model} to compare our methods and the SOTA baselines. The query asks about the time of the opening ceremonies for the 2018 Olympic Games, only THYME ranks the ground-truth table to the top. ``2018'', ``Olympics'', ``opening'', and ``ceremony'' occur in the titles and/or cells while ``when'' is semantically matched with ``Date'' and ``Time'' in the relevant table. In contrast, the strongest baselines BIBERT-SPLADE, and SSDR$_{im}$ rank the table that matches all the important words in the query while neglecting the semantic matching to ``when for 2018'' that needs date or time to answer. It shows that THYME has successfully learned field-aware hybrid matching. 
\section{Conclusion}
In this work, we propose a hybrid retriever based on the observation that table cells could prefer local matching of detailed information which differs from unstructured text such as table titles. We incorporate both dense and sparse representations to better suit the matching needs at smaller granularities. We tailor the representations for tables and propose a table-specific pooling method as well as a hybrid training technique. Experimental results show that our proposed method can adaptively balance the semantic and lexical matching requirements among the table fields. 
Tables, along with formats like text, HTML, and PDF, collectively constitute a vast amount of real-world data. Our proposed framework provides an idea for structured data retrieval represented by tables. Whether our proposed retriever works for other formats of structured data remains to be explored.
\section{Limitations}
This paper investigates the differences in matching across different fields in a table during the retrieval. However, due to resource limitations, no experiments were conducted using an LLM-based retriever, such as LLM2Vec~\citep{behnamghader2024llm2veclargelanguagemodels}. However, our optimization objective is orthogonal to that of these models, and their performance improvements can be combined. Additionally, tables are stored not only as text but also as images. Our proposed method is unsuitable for multimodal table retrieval. Future work will address these limitations.
\bibliography{acl_latex}
\newpage
\appendix
\section{Details of Baselines}\label{baseline_detail}
\textbf{Sparse Retrievers:}
\begin{itemize}[leftmargin=*]
    \item \textbf{BM25~\citep{robertson1994some}} is a well-known retrieval method that estimates the relevance of documents to a user query based on bag-of-words representations and exact term matching.
    \item \textbf{SPLADE~\citep{formal2021splade}} is a sparse retriever based on BERT and one of the backbones of our model. It maps a query or document to a vector of the vocabulary size, where each dimension corresponds to the probability of a term.
\end{itemize}
\textbf{Dense Retrievers:}
\begin{itemize}[leftmargin=*]
    \item \textbf{BIBERT~\citep{bibert-pretrain}} is a standard dense retriever based on BERT. It is also one of the backbones of our model. The hidden state of $[CLS]$ for a query and a document from BERT is used to estimate the relevance score.
    \item \textbf{DPR~\citep{wang-etal-2022-table}} demonstrates that the retriever that has been fine-tuned with a text corpus also can improve its performance in table retrieval.
    \item \textbf{TAPAS~\citep{Herzig_2020}} utilizes distinct types of embeddings like row and column embeddings to represent structure. It is also pre-trained on a large amount of tabular data and fine-tuned on cell, row, and column-level tasks. It is a universal table encoder that is widely used in table-related tasks.
    \item \textbf{DTR~\citep{herzig2021open}} uses TAPAS as the encoder and has been fine-tuned with relevant data of tables and queries. 
    \item \textbf{SSDR$_{im}$~\citep{jin2023enhancing}} is the state-of-the-art (SOTA) table retriever, which extracts the vectors of nouns to represent the query. For tables, it constructs representations of rows and columns by pooling, a part of which is sampled as the representation of tables. 

\end{itemize}
\textbf{Hybrid Retrievers}:
\begin{itemize}[leftmargin=*]
    \item \textbf{BIBERT+BM25} is a hybrid retrieval method that obtains the relevance score by directly adding the scores of semantic matching based on BIBERT and exact matching from BM25.
    \item \textbf{BIBERT+SPLADE} is a simple fusion of BIBERT and SPLADE. The outputs of both are used to estimate semantic matching and lexical matching respectively. Similar to BIBERT+BM25, the relevance score comes from the sum of the semantic matching score and lexical matching score.
    \item \textbf{DHR~\citep{lin2023aggretriever}} densifies the sparse representation and concatenates it with the dense representation to construct a single representation. It is compatible with most retrieval frameworks.
\end{itemize}




\end{document}